# Exploring the cylindrical photo-bending shape in polydomain nematic glass

Chen Xuan, Changwei Xu, Yongzhong Huo[1*]

Department of Mechanics and Engineering Science, Fudan University, Shanghai 200433, China

**Abstract**

This paper explores different photo-bending shapes in polydomain nematic glass. The motivation is to explain the phenomenon in experiment [1] under polarized light in which a nematic film curls into an circular arc, like part of a cylindrical surface. Polarized light triggers photo-isomerization and therefore makes liquid crystals (LCs) contract along their directors. We apply the Sachs limit to homogenize the deformation of polydomain LC glass. Photo-strain can be either contraction or expansion through the material. Bending shapes can be anticlastic, bowl-shaped and cylindrical affected by Poisson ratio and illumination intensity. An explanation for the cylindrical bend and ways to observe other shapes are given in a parameter plane.

*Keywords: liquid crystal, polydomain, homogenization, photo-bend*

## 1. Introduction

Liquid crystal polymers (LCPs) deform in an anisotropic way related to their directors **n** [1]-[8]. Nematic LCs in the glass state are stiff, their deformation often small, with hardly any director rotation. Photochromic nematic glass is nematic glass doped with photochromic moieties such as azo-dyes. Under polarized ultraviolet light (PUV) with a suitable wavelength, a nematic glass sample will contract along **n** and bend toward the light through a *trans-cis* photo-isomerization process [4]. The experiment on polydomain nematic glass [1] states that the sample is contracted along polarization and bent toward light into a partial cylindrical surface. Many theories study photo-deformation [4][10][11][20][24]-[27], of which [10] and [11] use finite element simulations to reproduce experiment [1]. [21]-[23] propose a polarization dependent photo-isomerization theory and study how PUV triggers reorientation of nematic directors in polydomain LC elastomers to reduce elastic energy. However, the exact deformation components and their distributions through the material have not been fully explored yet. So is the cylindrical bending shape lack of a convincing explanation. A good understanding of this phenomenon might reveal more insightful physics in photo-deformation.

When PUV light is illuminated onto a photochromic nematic film, a *trans-cis* photo-isomerization happens throughout the sample. One would believe that a polydomain sample should contract along polarization and expand in other directions. We will reveal that due to the interplay of domains in different orientations, this is not necessarily the case. Furthermore, as light attenuates through propagation, the *trans-cis* transition has a gradient distribution through film thickness. This leads people to believe intuitively that photo-strain is maximum contraction at illumination surface, minimum contraction at the back of the film, and so the film will bend towards the light with no doubt. Together with the popular assumption of contraction along polarization and expansion in lateral directions, one would believe an anticlastic bending shape should be observed. But experiment [1] shows exactly a non-anticlastic shape, for which hardly any explanation is given thus far. Our model reveals that pure decaying photo-contraction distribution and anticlastic bend only hold true for monodomains samples. We will show that polydomain photo-strain distributions can be contraction, expansion and can shift through propagation. Based on the distributions of photo-strain, different bending modes: anticlastic, bowl-shaped, cylindrical are predicted. From these bending modes we will find somewhere for experiment [1] to fall in, a possible explanation.

---

[1] Corresponding author: yzhuo@fudan.edu.cn

One would worry about the complicated microstructures in polydomain materials. In this paper, we use homogenization method to compute polydomain deformation. We take the Sachs limit ---- a lower energy bound compared to the upper bound in [22] ---- polydomain deformation is the average value of monodomain deformation within a representative volume element (RVE). This helps to shift our attention from tedious details of deformation in each domain to overall effective deformation of the polydomain material. This approach is a highly simplified model but we will show that it is good to reveal some rich properties in polydomain deformation.

## 2. Homogenization of photo response in polydomain nematic glass

Micromechanics is widely applied when considering effects of microstructures on materials. These physical and mechanical properties include modulus, density, heat capacity, electric resistance *etc*. Due to complexity of microstructures, simplification and approximation is necessary. People use homogenization to study approximate physical and mechanical properties of heterogeneous materials. Homogenization is to find a homogeneous material which is equivalent to the heterogeneous one in the sense of energy. The properties of the homogeneous material become effective ones. For polycrystal or polydomain composite materials, homogenization of the physical quantities is also a basic approach [12]-[15]. In this case, the polydomain strain $\boldsymbol{\varepsilon}^p$ for nematic glass can be obtained by adding up the strain $\boldsymbol{\varepsilon}^m(\mathbf{r})$ in respective domains

$$\boldsymbol{\varepsilon}^p = \int_\Omega \boldsymbol{\varepsilon}^m(\mathbf{r})\rho(\mathbf{r})dV = V<\boldsymbol{\varepsilon}^m(\mathbf{r})>, \tag{1}$$

where the distribution function of the domains $\rho(\mathbf{r})$ is normalized. This equation is valid only when the deformation is small. Likewise, the polydomain stress $\boldsymbol{\sigma}^p$ for nematic glass can be obtained by adding up the stress $\boldsymbol{\sigma}^m(\mathbf{r})$ in respective domains

$$\boldsymbol{\sigma}^p = \int_\Omega \boldsymbol{\sigma}^m(\mathbf{r})\rho(\mathbf{r})dV = V<\boldsymbol{\sigma}^m(\mathbf{r})>. \tag{2}$$

The above overall effective stress can also be regarded as a linear transformation of stresses in respective domains or be written inversely as [17]

$$\boldsymbol{\sigma}^m(\mathbf{r}) = \mathbf{A} \cdot <\boldsymbol{\sigma}^m(\mathbf{r})>. \tag{3}$$

$\mathbf{A}$ related to the linear transformation is determined by the distribution of the domains.

To apply homogenization, one starts from an RVE that is large enough compared with a microscopic scale but small enough compared with a macroscopic scale. Here, an RVE includes many domains in nematic glass. The strain caused by impurity, thermal strain for thermo-elastic materials and plastic strain for elasto-plastic materials are called eigen-strain. Apart from the calculation of strain, the elastic constants of polydomain materials should be modified by homogenization. When deformation is rather significant, the elastic moduli should be delicately calculated due to different deformation in respective domains [16]-[19]. For general materials with impurity or for thermo-elastic, elasto-plastic materials, the deformation (when infinitesimal) is described by a total strain which is an elastic strain

plus an eigen-strain $\boldsymbol{\varepsilon}_{ph}^m$

$$\boldsymbol{\varepsilon}^m = \mathbf{S}^m : \boldsymbol{\sigma}^m + \boldsymbol{\varepsilon}_{ph}^m, \tag{4}$$

where $\mathbf{S}^m$ is the compliance tensor in a domain. Note that the above decomposition of the total strain only holds true when the strain is small. Taking an average value within the whole region gives

$$<\boldsymbol{\varepsilon}^m> = <\mathbf{S}^m \cdot \mathbf{A}> :<\boldsymbol{\sigma}^m(\mathbf{r})> + <\boldsymbol{\varepsilon}_{ph}^m>. \tag{5}$$

The overall strain-stress relation for polydomain nematic glass holds

$$\boldsymbol{\varepsilon}^p = \mathbf{S}^p : \boldsymbol{\sigma}^p + \boldsymbol{\varepsilon}_{ph}^p, \tag{6}$$

where $\mathbf{S}^p$ is the effective compliance tensor for polydomain nematic glass yet to be determined. Combining the above two expressions leads to

$$\mathbf{S}^p = \int_\Omega \mathbf{S}^m \cdot \mathbf{A} \rho(\mathbf{r}) dV \tag{7}$$

and

$$\boldsymbol{\varepsilon}_{ph}^p = \int_\Omega \boldsymbol{\varepsilon}_{ph}^m(\mathbf{r}) \rho(\mathbf{r}) dV. \tag{8}$$

The effective compliance tensor or elasticity tensor for polydomain nematic glass could be computed theoretically given the distribution of the domains and the compliance tensors in the respective domains. Likewise, the effective eigen-strain for polydomain nematic glass could be computed theoretically given the distribution of the domains and the eigen-strain in the respective domains.

However, when deformation is small, the elastic moduli could be roughly regarded as uniform throughout the sample. We would consider the case in which light induced strain for nematic glass is small, since nematic glass itself is relatively stiff. We assume that the elastic moduli are uniform and remain the same values before and after light illumination. Therefore, the only thing we care about in this paper is the effective photo-strain rather than the effective elastic moduli for polydomain nematic glass.

Two limits can be used when doing further approximation [28][29]. One is the Taylor limit in which all domains undergo the same deformation but this limit does not require stress balance. Therefore it can be an upper bound on energy since the system requires external energy and force on top of what is exactly needed to keep in equilibrium. This limit is what [22] uses to calculate polydomain photo-strain. The other limit is the Sachs limit in which all domains suffer the same stress but deformation is lack of compatibility. This approach sets a lower bound on energy since it allows test strain fields which are even incompatible on the domain boundaries, which enlarges the confine of allowed test strain solutions. This limit is what we use in this paper. As photo-strain in nematic glass is usually small, we could ignore the stress in domain-domain interactions caused by light in the first instance, though more careful calculation should be done in future studies. Hooke's Law can serve as a good approximation in this limit and hence the elastic strain could be ignored. So the polydomain real strain is approximately the polydomain photo-strain

$$\boldsymbol{\varepsilon}^p \approx \boldsymbol{\varepsilon}_{ph}^p = \int_\Omega \boldsymbol{\varepsilon}_{ph}^m(\mathbf{r}) \rho(\mathbf{r}) dV. \tag{9}$$

The eigen-strain herein is the effective polydomain photo-strain. But this eigen-strain is different from general thermal strain or plastic strain. Light has a wave vector $\mathbf{k}$. The photo-strain is dependent on $\mathbf{k}$. For diffused ultraviolet light (DUV), one can compute the polydomain photo-strain $\boldsymbol{\varepsilon}^p(\mathbf{k})$ according to the photo-isomerization [21]-[23], given $\boldsymbol{\varepsilon}^m_{ph}(\mathbf{r};\mathbf{k})$ and the domain distribution $\rho(\mathbf{r})$. For PUV, light has an extra direction, namely the polarization direction $e$. According to the photo-isomerization [21]-[23], one can compute the polydomain photo-strain $\boldsymbol{\varepsilon}^p(\mathbf{k},\mathbf{e})$, given $\boldsymbol{\varepsilon}^m_{ph}(\mathbf{r};\mathbf{k},\mathbf{e})$ and the domain distribution $\rho(\mathbf{r})$.

### 3. Polarization dependent surface photo-strain of mono- and polydomain nematic glass

*3.1 Polarization dependent surface photo-strain in monodomain nematic glass*

We consider a monodomain photochromic nematic glass with azobenzene illuminated normal to the surface by linearly PUV light with intensity $I_0$, causing *trans-cis* reaction alongside with thermal back reaction. The energy is lower in the absence of light when the azobenzene molecules are in the rod like *trans* state. The energy is lower in the presence of light when the azobenzene molecules are in the curling *cis* state. The photochromic azobenzene molecules are converted from the rod like *trans* state to the curling *cis* state known as photo-isomerization. The population of *cis*-isomers or called *cis* fraction $n_c$ under light intensity $I$ is determined by the competition of the bare *trans-cis* transition and the *cis-trans* back reaction. The $n_c$ satisfies the following differential equation [21]-[23]

$$\frac{\partial n_c}{\partial t} = \eta_0 \hat{\eta} I (1-n_c) - \tau_{ct}^{-1} n_c, \tag{10}$$

where $\eta_0$ is a positive constant and $\tau_{ct} = \exp(\Delta/k_B T)$ is the characteristic time for the *cis-trans* thermal back transition. $\hat{\eta}$ is the dimensionless absorption coefficient with polarization dependence. It introduces the average effect of the angles between the polarization and the nematic molecule directions

$$\hat{\eta}(\psi;Q) = 1 + 2Q P_2(\cos\psi) = 1 + Q(3\cos^2\psi - 1), \tag{11}$$

where $\cos\psi = \mathbf{n}\cdot\mathbf{e}$, $e$ and $n$ are respectively the polarization direction of the PUV and the nematic director, and $Q$ is called either the degree of orientation or order parameter. $P_2(\cos\psi) = (3\cos^2\psi - 1)/2$ is the second Legendre polynomial. $\psi$ is the relative angle between the director $n$ and the polarization $e$, and will be simply called polarization angle in later contexts.

Though dynamics of *trans-cis* isomerization could be interesting both experimentally and theoretically, we study photo-stationary state and its effects on bending mechanics in this paper for theoretical convenience. This approach is not lack of experimental merit, because photo-stationary state can always be expected as one waits for long enough, whereas instantaneous observation might take more efforts. So when the isomerization process progresses long enough, $n_c$ comes to a steady state solution by letting $\partial n_c / \partial t$ be zero and satisfies

$$n_c = \frac{\hat{\tau}_{ct}\hat{\eta}(\psi;Q)i_0 i}{1+\hat{\tau}_{ct}\hat{\eta}(\psi;Q)i_0 i}, \tag{12}$$

where $\hat{\tau}_{ct}(T) = \tau_{ct}(T)/\tau_{ct}(T_0)$, $T_0$ is a reference temperature, $i_0 = \tau_{ct}(T_0)\eta_0 I_0$ is a dimensionless surface light intensity, $I_0$ is the light intensity at the illumination surface and $i = I/I_0$. In this section $i = 1$ since we only consider the photo-isomerization at the illumination surface.

The photochromic azobenene molecules are crosslinked onto the nematic glass. As photo-isomerization changes the microscopic azobenzene molecule configuration, the molecule order of nematic glass is partly destroyed by the PUV light through crosslinking. The order-mechanical coupling triggers a macroscopic deformation of the nematic glass sample, specifically written as the following photo-strain tensor [25][27]

$$\boldsymbol{\varepsilon}^m_{ph}(\mathbf{n},\mathbf{e}) = \varepsilon^m_{\|\mathbf{n}}(\mathbf{n}\cdot\mathbf{e})[-\nu^m \mathbf{I} + (1+\nu^m)\mathbf{n}\otimes\mathbf{n}]. \tag{13}$$

Namely, the nematic glass sample contracts axially along the orientation $\mathbf{n}$ by a magnitude of $-\varepsilon^m_{\|\mathbf{n}} > 0$, which depends on $\cos\psi = \mathbf{n}\cdot\mathbf{e}$, and expands laterally in the two orthogonal directions normal to $\mathbf{n}$ by $\nu^m$ times, where $\nu^m$ is the photo-Poisson ratio of the nematic glass (Fig. 1a). As photo-strain is small in the situation considered, the eigen-contraction along the nematic director can be approximated as linearly proportional to the *cis* fraction [25]

$$\varepsilon^m_{\|\mathbf{n}}(\psi) = -\gamma_\varepsilon n_c(\psi). \tag{14}$$

Combining (12) and (14), $\varepsilon^m_{\|\mathbf{n}}$, the photo-strain along the orientation direction $\mathbf{n}$, is shown as a function of the polarization angle $\psi$ in Fig. 1b. The photo-strain $\varepsilon^m_{\|\mathbf{n}}$ itself is negative and the magnitude of the photo-contraction $-\varepsilon^m_{\|\mathbf{n}}$ decreases with the polarization angle $\psi$. The fact that the photo-contraction originated from the *trans-cis* transition is the best when $\mathbf{e}\|\mathbf{n}$ and the worst when $\mathbf{e}\perp\mathbf{n}$, is not surprising as one can derive from (10), (11) and (14).

### 3.2 Polarization dependent surface photo-strain in polydomain nematic glass

We consider a thin polydomain nematic film illuminated by linearly PUV with intensity $I_0$, hence the same light illumination as in the last subsection but with different domains aligning in the nematic glass. Fig. 2 is a schematic diagram of a microelement extracted from a polydomain nematic film. This microelement contains many domains randomly distributed in the film plane. The entire nematic film is a planar extension of such microelements to ensure that the directors have no positional but orientational differences and that light is uniform falling on all domains. The domains align parallel to the illumination surface and their distribution is totally random ---- there is no macroscopic orientational preference for the overall film. As stated in the previous subsection, we assume an infinitesimal deformation limit and therefore stress in all domains is negligibly small, hence the Sachs limit (or called independent domain model) is applied. Director rotation is not considered here for relatively harder glass. Now $\psi$ is the relative angle between polarization and the director in an individual domain, and of course varies from domain to domain. We choose a distribution type: transverse isotropic distribution of domains, which is commonly seen in experiments, as an example. The so-called transverse isotropic distribution of domains is defined that all domain orientations lie within the plane of the film which is perpendicular to light illumination (Fig. 2). When a film is thin, directors often align parallel to the film plane due to many of the sample preparation methods. Based on our Sachs homogenization method (9), equivalent polydomain photo-strain tensor $\boldsymbol{\varepsilon}^p$ can be homogenized as the orientational average of photo-strain tensors $\boldsymbol{\varepsilon}^m_{ph}(\nu^m,\psi)$ in (13) over all domains

$$\boldsymbol{\varepsilon}^p(\nu^m) = \frac{1}{\pi}\int_0^\pi \boldsymbol{\varepsilon}^m_{ph}(\nu^m,\psi)d\psi . \qquad (15)$$

The orientation of the director in a domain is expressed as $\mathbf{n} = (\cos\psi, \sin\psi, 0)^T$ in the Cartesian coordinates ($x$, $y$, $z$), where $\psi \in [0,\pi)$ (Fig. 2). The polarization direction $e$ is assumed to be along the $x$ direction. It is not hard to show with the aid of (13) that the polydomain photo-strain tensor is diagonal, $\boldsymbol{\varepsilon}^p = diag\left(\varepsilon^p_{\|\mathbf{e}}, \varepsilon^p_{\perp\mathbf{e}}, \varepsilon^p_z\right)$, in the current Cartesian coordinate system. The three photo-strain components can be written as

$$\varepsilon^p_{\|\mathbf{e}}(\nu^m) = \frac{2}{\pi}\int_0^{\pi/2}[-\nu^m + (1+\nu^m)\cos^2\psi]\varepsilon^m_{\|\mathbf{n}}(\psi)d\psi , \qquad (16)$$

$$\varepsilon^p_{\perp\mathbf{e}}(\nu^m) = \frac{2}{\pi}\int_0^{\pi/2}[-\nu^m + (1+\nu^m)\sin^2\psi]\varepsilon^m_{\|\mathbf{n}}(\psi)d\psi , \qquad (17)$$

$$\varepsilon^p_z(\nu^m) = -\frac{2}{\pi}\nu^m\int_0^{\pi/2}\varepsilon^m_{\|\mathbf{n}}(\psi)d\psi . \qquad (18)$$

It is apparent that since photo-strain $\varepsilon^m_{\|\mathbf{n}}$ is contraction and therefore negative, $\varepsilon^p_z$ is always positive for $\forall \nu^m > 0$, hence a trivial expansion through the light propagation direction. With (12), (14), (16), (17), one reduces $\varepsilon^p_{\|\mathbf{e}}, \varepsilon^p_{\perp\mathbf{e}}$ to

$$\frac{\varepsilon^p_{\|\mathbf{e}}(\nu^m)}{\gamma_\varepsilon} = \frac{1+\nu^m}{3i_0 Q} + \frac{\nu^m - 1}{2} - \frac{1}{3i_0 Q}\frac{i_0 Q(2\nu^m - 1) + (1+\nu^m)(1+i_0)}{\sqrt{[1+i_0(1+2Q)][1+i_0(1-Q)]}} , \qquad (19)$$

$$\frac{\varepsilon^p_{\perp\mathbf{e}}(\nu^m)}{\gamma_\varepsilon} = -\frac{1+\nu^m}{3i_0 Q} + \frac{\nu^m - 1}{2} - \frac{1}{3i_0 Q}\frac{i_0 Q(\nu^m - 2) - (1+\nu^m)(1+i_0)}{\sqrt{[1+i_0(1+2Q)][1+i_0(1-Q)]}} . \qquad (20)$$

For commonly observed incompressible nematics, one could check the incompressibility of the polydomain samples by computing $tr(\boldsymbol{\varepsilon}^p) = 0$ while setting $\nu^m = 1/2$. Yet that "incompressibility" does not bring about an effective polydomain Poisson ratio $\nu^p = 1/2$! One would see in later paragraphs that $v^p \triangleq -\varepsilon_{\perp \mathbf{e}}^p / \varepsilon_{\| \mathbf{e}}^p$ could be positive, zero or negative. Incompressible (when $v^m = 1/2$) polydomain photo-strain components are simplified as (skipping the trivial $z$ component) from (16) and (17)

$$\varepsilon_{\| \mathbf{e}}^p(1/2) = \frac{2}{\pi} \int_0^{\pi/2} P_2(\cos^2 \psi) \varepsilon_{\| \mathbf{n}}^m(\psi) d\psi, \tag{21}$$

$$\varepsilon_{\perp \mathbf{e}}^p(1/2) = \frac{2}{\pi} \int_0^{\pi/2} [\frac{1}{2} - P_2(\cos^2 \psi)] \varepsilon_{\| \mathbf{n}}^m(\psi) d\psi. \tag{22}$$

One reduces them to

$$\frac{\varepsilon_{\| \mathbf{e}}^p(1/2)}{\gamma_\varepsilon} = \frac{1}{2i_0 Q}(1 - \frac{1+i_0}{\sqrt{[1+i_0(1+2Q)][1+i_0(1-Q)]}}) - \frac{1}{4}, \tag{23}$$

$$\frac{\varepsilon_{\perp \mathbf{e}}^p(1/2)}{\gamma_\varepsilon} = \frac{1}{2i_0 Q}(-1 + \frac{1+i_0(1+Q)}{\sqrt{[1+i_0(1+2Q)][1+i_0(1-Q)]}}) - \frac{1}{4}. \tag{24}$$

Photomechanical properties of nematic glass can be reflected via investigating these photo-strain components. For incompressible nematic glass namely $v^m = 1/2$, it is not hard to show from (23) that $\varepsilon_{\| \mathbf{e}}^p(1/2) < 0$ for $\forall i_0 > 0, \forall Q \in (0,1)$. That means that the photo-strain along the polarization *e* is always contraction at the illumination surface, which is in line with our intuition and experiment [1]. Interestingly, $\varepsilon_{\perp \mathbf{e}}^p(1/2)$ changes its sign according to the parameters $i_0$, $Q$. We discover that there exists a critical order parameter $Q^0(i_0)$ for $\varepsilon_{\perp \mathbf{e}}^p(1/2) = 0$, *i.e.* When $Q < Q^0(i_0)$, $\varepsilon_{\perp \mathbf{e}}^p(1/2) < 0$; when $Q > Q^0(i_0)$, $\varepsilon_{\perp \mathbf{e}}^p(1/2) > 0$. It is not hard to derive from (24) that $Q^0|_{i_0 \to 0} = 4/7$, in the Beer weak light limit. As $\varepsilon_{\perp \mathbf{e}}^p(1/2)$ can be either expansion or contraction while $\varepsilon_{\| \mathbf{e}}^p(1/2) < 0$, the effective polydomain photo-Poisson ratio $v^p$ does not hold a definite sign, though incompressibility means Poisson ratio equals 1/2 in conventional materials.

$\varepsilon_{\| \mathbf{e}}^p$ and $\varepsilon_{\perp \mathbf{e}}^p$ are a result of the interplay between the polarization dependence (11) and monodomain photo-Poisson ratio $v^m$. In an individual domain deformation are axial $\varepsilon_{\| \mathbf{n}}^m$ along its director and lateral $-v^m \varepsilon_{\| \mathbf{n}}^m$ normal to its director (13). According to the polarization dependence (11), both are big when the director is at a small angle to polarization, and vice versa. To simplify analysis, one can always imagine that domains were either exactly along *x* or *y*. Then, $\varepsilon_{\| \mathbf{e}}^p$ depends on axial contraction along *x* of *x* domains and lateral expansion along *x* ($v^m$ times axial contraction along *y*) of

$y$ domains. For relatively small $v^m$, 1/2 inclusive, $\varepsilon^p_{\|\mathbf{e}} < 0$ will be obtained, since axial contraction $\varepsilon^m_{\|\mathbf{n}}$ of $x$ domains is bigger than axial $\varepsilon^m_{\|\mathbf{n}}$ of $y$ domains, see Fig. 1b. In contrast, $\varepsilon^p_{\perp\mathbf{e}}$ depends on axial contraction along $y$ of $y$ domains and lateral expansion along $y$ ($v^m$ times axial contraction along $x$) of $x$ domains. For small $v^m$, it is impossible to judge if $x$ or $y$ domains dominates the strain, since the axial contraction $\varepsilon^m_{\|\mathbf{n}}$ of $y$ domains is smaller than $\varepsilon^m_{\|\mathbf{n}}$ of $x$ domains. For big $v^m$, the above analysis gives a contrary result (refer to Fig. 5 for $z=0$ in the next section). The above explains why not always photo-contraction along $e$ and photo-expansion normal to $e$.

In addition to the indefinite signs of $\varepsilon^p_{\|\mathbf{e}}$ and $\varepsilon^p_{\perp\mathbf{e}}$, we have discovered when $|\varepsilon^p_{\|\mathbf{e}}|$ could be predominantly more important than $|\varepsilon^p_{\perp\mathbf{e}}|$ through adjusting $i_0$ and $v^m$. Without loss of generality we set $Q = 0.453 < Q^0\big|_{i_0 \to 0} = 4/7$ and the rest is analogous. The parameter area for $|\varepsilon^p_{\perp\mathbf{e}} / \varepsilon^p_{\|\mathbf{e}}| < 20\%$ and the curve for $\varepsilon^p_{\perp\mathbf{e}} = 0$ in the $(i_0, v^m)$ plane are shown in Fig. 3. If $|\varepsilon^p_{\perp\mathbf{e}}|$ is, say, only 20% of $|\varepsilon^p_{\|\mathbf{e}}|$, lateral deformation normal to $e$ may well be unobserved in experiments compared to that along $e$. If basically $\varepsilon^p_{\|\mathbf{e}}$ causes bend, $x$ domains are more bent than $y$ ones. This curvature difference along $x$ and $y$ results that the bending stiffness along $y$ becomes significantly bigger making $\varepsilon^p_{\perp\mathbf{e}}$ harder to grow were one to continue shining light as it bends (see also Fig. 7a). We predict that in large deflection or even curling $\varepsilon^p_{\perp\mathbf{e}}$ might not be able to develop by continuous illumination due to the bent configuration. This can be a possible explanation for the cylindrical-shape bend observed in [1], where $|\varepsilon^p_{\perp\mathbf{e}}|$ looks much smaller than $|\varepsilon^p_{\|\mathbf{e}}|$. Similar analysis for curvature ratio will be seen in the next section. But on the other hand we point out that cylindrical bend in [1] is always observable, as one choose parameters outside the $|\varepsilon^p_{\perp\mathbf{e}} / \varepsilon^p_{\|\mathbf{e}}| < 20\%$ region in Fig. 3.

In this section, we briefly introduce the polarization dependent photo-isomerization and its effect on the surface deformation of nematic glass. The punchline of the polarization dependence is that *trans-cis* transition is maximum when polarization is parallel to nematic orientation and is minimum when normal. In polydomain samples, whether contraction or expansion at the surface strongly depends on Poisson ratio. When $v^m = 1/2$, $\varepsilon^p_{\|\mathbf{e}} < 0$ ---- definite contraction along $x$ ---- but $\varepsilon^p_{\perp\mathbf{e}}$ depends on intensity and order parameter. Through $|\varepsilon^p_{\perp\mathbf{e}} / \varepsilon^p_{\|\mathbf{e}}|$ we give a possible explanation for the cylindrical-shape bend observed in [1].

## 4. Polarization dependent photo-strain distribution and photo-bend in polydomain nematic glass

*4.1 Polarization dependent photo-strain distribution through propagation*

So far all we have studied is photo-strain at illumination surface. In this section, our focus is on a specific structure, namely a nematic glass film. We will have to consider light decay through the LC medium due to light propagation. Then photo-strain varies normal through the film since photo-isomerization depends on light intensity entered. Gradient distribution of photo-strain brings about photo-bend, which we measure through curvature.

We consider a nematic film that is much thinner than the optic characteristic absorption length ($h \ll d$) as in [21]-[23] such that hardly any change of polarization or depolarization could happen ---- polarization would always remain constant from where it enters to where it leaves through film thickness direction. This assumption would be deficient in accuracy, especially when the film thickness is comparable to the depolarization length. We still use this approximation not only out of theoretical convenience but also because in what follows our results do give some telling explanation for experiments. Therefore light absorption is described by the modified Lambert-Beer's Law in which light intensity $I$ decays in an exponential manner with polarization dependence [21]-[23]:

$$\frac{\partial I}{\partial z} = -\gamma_t \eta_0 \hat{\eta} I (1 - n_c), \tag{25}$$

where $\gamma_t$ is a positive material constant and the optic absorption characteristic length is defined as $d = 1/\gamma_t \eta_0$. Then one combines (10) with (25) to solve $I$ and $n_c$

$$\frac{I}{I_0} = \frac{W(i_0 \hat{\eta} e^{(i_0 - z/d)\hat{\eta}})}{i_0 \hat{\eta}}, \text{ with } \hat{\eta} = \hat{\eta}(\psi; Q), \tag{26}$$

$$n_c = \frac{W(i_0 \hat{\eta} e^{(i_0 - z/d)\hat{\eta}})}{1 + W(i_0 \hat{\eta} e^{(i_0 - z/d)\hat{\eta}})}, \text{ with } \hat{\eta} = \hat{\eta}(\psi; Q). \tag{27}$$

With the linear relation between $\varepsilon_{\parallel \mathbf{n}}^m$ and $n_c$ (14), one could obtain the distribution of monodomain photo-strain through an LC medium, shown in Fig. 4. $|\varepsilon_{\parallel \mathbf{n}}^m|$ is small for large polarization angles $\psi \in [0, \pi/2]$ near the illumination surface ($z=0$). At larger depths, it is the opposite, namely, large $|\varepsilon_{\parallel \mathbf{n}}^m|$ for large polarization angles. That is because parallel illumination facilitates both photon absorption and *trans* depletion. Although parallel illumination depletes *trans* the most at the surface, light is most absorbed and reduced, which in turn acts against *trans* depletion. Despite a high *cis* fraction at the surface, $n_c$ will eventually drop once light needed for triggering isomerization is depleted through decay. More complicated polydomain photo-strain distribution is based on this mechanism in monodomain LCs.

Polydomain photo-strain tensor is still (15). So do (16)-(18) satisfy the polydomain photo-strain components when light absorption is considered. One gets polydomain strain components through inserting (27) into (16)-(18). As the monodomain photo-strain $\varepsilon_{\|\mathbf{n}}^m(z)$ decays monotonically through propagation (Fig. 4), one might expect a similar distribution of the polydomain photo-strain as well. But the situation is far more complex than it seems. Fig. 5 shows how $\varepsilon_{\|\mathbf{e}}^p(z)$ and $\varepsilon_{\perp\mathbf{e}}^p(z)$ are distributed through propagation, according to (16) and (17). For nematic liquid crystal polymers, networks or glass, the Poisson ratio is reported to range from 0 to 2 [4].

First look at the extreme but simpler case when $v^m = 0$, where there would be pure axial contraction with no lateral expansion within all individual domains. Both $\varepsilon_{\|\mathbf{e}}^p(z)$ and $\varepsilon_{\perp\mathbf{e}}^p(z)$ are contraction and decay through $z$ without surprise (Fig. 5). $\varepsilon_{\|\mathbf{e}}^p(z)$ is larger than $\varepsilon_{\perp\mathbf{e}}^p(z)$ in magnitude because $x$ domains align with $e$ and contract axially more than $y$ domains do.

For $v^m > 0$, domains do undergo lateral expansion while they contract axially. This brings about an interesting polydomain strain distribution that for small $v^m > 0$, $\varepsilon_{\|\mathbf{e}}^p(z)$ is contraction near the illumination surface but expansion below a critical depth (Fig. 5a). The coexistence of both contraction and expansion for $\varepsilon_{\|\mathbf{e}}^p(z)$ in its distribution through propagation runs a bit contrary to popular belief, though the expansion is much smaller than the contraction in magnitude. How those lateral expansion of domains comes into play can be understood by looking at Fig. 4. At the illumination surface, as $\varepsilon_{\|\mathbf{n}}^m(z)\big|_{\psi=0}$ is bigger than $\varepsilon_{\|\mathbf{n}}^m(z)\big|_{\psi=\pi/2}$ ---- axial contraction of $x$ domains is bigger than that of $y$ domains ---- axial contraction of $x$ domains then beats lateral expansion of $y$ domains for small $v^m$, which is just $v^m$ times the axial contraction of $y$ domains. Deep below the critical depth, as $\varepsilon_{\|\mathbf{n}}^m(z)\big|_{\psi=\pi/2}$ is bigger than $\varepsilon_{\|\mathbf{n}}^m(z)\big|_{\psi=0}$ ---- axial contraction of $y$ domains overtakes that of $x$ domains ---- such that lateral expansion of $y$ domains beats axial contraction of $x$ domains as well. This competition causes such a switchover in $\varepsilon_{\|\mathbf{e}}^p(z)$ distribution. But it is noteworthy that for a specific structure, the critical depth for $\varepsilon_{\|\mathbf{e}}^p(z)$ to be zero does not necessarily exist as it depends on the actual thickness of the structure.

Following the above analysis it is easy to understand that when $v^m$ is big enough: $\varepsilon_{\|\mathbf{e}}^p(z) > 0$ for all $z$; $\varepsilon_{\perp\mathbf{e}}^p(z)$ is expansion near the surface and contraction deep below the surface. The curves for all other $v^m$ that are not shown in Fig. 5 are easy to infer thanks to the linear dependence of both $\varepsilon_{\|\mathbf{e}}^p(z)$ and $\varepsilon_{\perp\mathbf{e}}^p(z)$ on $v^m$, as seen in (16) and (17). Interestingly for all $v^m$, the magnitudes of

$\varepsilon_{\parallel\mathbf{e}}^{p}(z)$ and $\varepsilon_{\perp\mathbf{e}}^{p}(z)$ are much smaller than $\varepsilon_{\parallel\mathbf{n}}^{m}$, due to the interplay among different domains, also seen in [9].

In this subsection by considering light absorption, polydomain strain distribution through propagation is studied. Adjustable by Poisson ratio $\nu^m$, axial contraction along polarization and lateral expansion would not always be observed ---- $\varepsilon_{\parallel\mathbf{e}}^{p}(z)$ and $\varepsilon_{\perp\mathbf{e}}^{p}(z)$ could change signs through propagation. Based on the polarization dependence in individual domains, $\nu^m$ interacts with the photo-strain in a more complex manner.

*4.2 Polarization dependent photo-bend*

Gradient photo-strain leads to curvature. As the features of $\varepsilon_{\parallel\mathbf{e}}^{p}(z)$ and $\varepsilon_{\perp\mathbf{e}}^{p}(z)$ explored above are quite complex, curvature must be more involved. The small deflection photochromic plate model [27] is applied here to study bend because plate tilts in small deflection are still roughly perpendicular to illumination ---- otherwise light falling on a strongly deflected plate at skewed angles involves more optical complications. Effective overall elastic parameters can be viewed as isotropic were the isotropic domain distribution assumed (Fig. 2). Fig. 2 is the simplest case for a nematic film/plate ($a \times b \times h$) with free boundary conditions for all the four lateral edges. The free boundary conditions bring about a simple solution for the longitudinal deflection *w*. The corresponding dimensionless in-plane principal curvature along the *x* and *y* directions are [27]

$$\kappa_{\parallel\mathbf{e}}^{p} = -\frac{h}{12}\frac{\partial^2 w}{\partial x^2} = \frac{1}{h^2}\int_0^h \varepsilon_{\parallel\mathbf{e}}^{p}(z)(z-\frac{h}{2})dz, \quad \kappa_{\perp\mathbf{e}}^{p} = -\frac{h}{12}\frac{\partial^2 w}{\partial y^2} = \frac{1}{h^2}\int_0^h \varepsilon_{\perp\mathbf{e}}^{p}(z)(z-\frac{h}{2})dz \,. \quad (28)$$

If $\kappa_{\parallel\mathbf{e}}^{p}$ is positive, the film bends towards the light in the *x-z* plane; if $\kappa_{\perp\mathbf{e}}^{p}$ is positive, the film bends towards the light in the *y-z* plane; and vice versa.

In the first instance, we study photo-bend in incompressible nematic glass, namely $\nu^m = 1/2$. Recall the photo-strain distribution in Fig. 5a. $\varepsilon_{\parallel\mathbf{e}}^{p}(z)$ is contraction near the surface and decays through propagation. We investigate all parameters including changing order parameter, light intensity, and come to the conclusion the distribution curve of $\varepsilon_{\parallel\mathbf{e}}^{p}(z)$ remains qualitatively the same. Then it is not hard to find with the aid of (28) that $\kappa_{\parallel\mathbf{e}}^{p}$ is always positive for any positive *h* and $i_0$. That means that the film (when $\nu^m = 1/2$) will always bend towards the light in the *x-z* plane, which is in perfect agreement with experiment [1]. In addition $\kappa_{\parallel\mathbf{e}}^{p}$ is found to be a non-monotonic function of *h* in a similar fashion obtained in [24].

$\varepsilon_{\perp\mathbf{e}}^{p}(z)$ shows a different distribution in Fig. 5b ($\nu^m = 1/2$). If one changes the light intensity,

the distribution is sensitive to intensity as shown in Fig. 6a. The monotonicity of the $\varepsilon_{\perp\mathbf{e}}^{p}(z)$ curve changes with $i_0$. $\varepsilon_{\perp\mathbf{e}}^{p}(z)$ can be classified into two types: $\left|\varepsilon_{\perp\mathbf{e}}^{p}(z)\right|$ increases at the surface (type I) and decreases at the surface (type II). Under weak light namely $i_0 < i_0^c$, $\left|\varepsilon_{\perp\mathbf{e}}^{p}(z)\right|$ increases at the surface until some critical $z$ and it decreases to approach 0 (type I). When $i_0 = i_0^c$, $d\varepsilon_{\perp\mathbf{e}}^{p}(z)/dz\big|_{z=0} = 0$. When $i_0 > i_0^c$, $\left|\varepsilon_{\perp\mathbf{e}}^{p}(z)\right|$ decreases at the surface (type II). This change in the distribution of $\varepsilon_{\perp\mathbf{e}}^{p}(z)$ seems innocuous, but we show below that it plays an essential role in determining the sign of $\kappa_{\perp\mathbf{e}}^{p}$.

An easy way of revealing the connection between photo-strain and signs of curvature is through Taylor expansion with respect to $h$. When $h/d \ll 1$, it is not hard to show a relation between the curvature and photo-strain expanded to the first order of $h/d$ by manipulating (28), (16) and (17)

$$\kappa_{\|\mathbf{e}}^{p}\big|_{h\to 0} = \frac{h}{12}\frac{d\varepsilon_{\|\mathbf{e}}^{p}}{dz}\bigg|_{z=0}, \kappa_{\perp\mathbf{e}}^{p}\big|_{h\to 0} = \frac{h}{12}\frac{d\varepsilon_{\perp\mathbf{e}}^{p}}{dz}\bigg|_{z=0}. \tag{29}$$

It takes little effort to check that $\kappa_{\|\mathbf{e}}^{p}\big|_{h=0} = \kappa_{\perp\mathbf{e}}^{p}\big|_{h=0} = 0$, for when $h$ approaches 0 the photo-strain appears too homogeneous through the film to bend. But due to the interesting distribution of $\varepsilon_{\perp\mathbf{e}}^{p}(z)$ (Fig. 6a), $\kappa_{\perp\mathbf{e}}^{p}(h)$ can change its sign unlike $\kappa_{\|\mathbf{e}}^{p}(h)$: (a) When $i_0 < i_0^c$, $\varepsilon_{\perp\mathbf{e}}^{p}(z)$ decreases at the surface (type I), so $\kappa_{\perp\mathbf{e}}^{p}\big|_{h\to 0} < 0$ indicating that $\kappa_{\perp\mathbf{e}}^{p}(h)$ decreases from 0. However, due to the overall distribution of $\kappa_{\perp\mathbf{e}}^{p}(h)$ throughout the whole sample, $\kappa_{\perp\mathbf{e}}^{p}(h) > 0$ when $h$ is big enough. So there must be at least one zero curvature point for some $h>0$; (b) When $i_0 = i_0^c$, as $d\varepsilon_{\perp\mathbf{e}}^{p}(z)/dz\big|_{z=0} = 0$, so $\kappa_{\perp\mathbf{e}}^{p}\big|_{h\to 0} = 0$, and $\kappa_{\perp\mathbf{e}}^{p}(h)$ will eventually increase to be positive as $h$ increases; (c) When $i_0$ continues to increase (type II), $\kappa_{\perp\mathbf{e}}^{p}(h)$ starts to increase from zero but soon decreases to negative values and increases back to positive values before decaying to 0. In this case there would be two zero curvature points for $h>0$; (d) When $i_0$ is big enough, the above two zero curvature points coalesce to one point and eventually no zero point for $h>0$ any more. Due to the nonlinearity of the curvature $\kappa_{\perp\mathbf{e}}^{p}(h)$ with respect to $h$, it is possible that there exist at most two critical $h_0$ for $\kappa_{\perp\mathbf{e}}^{p}(h)$ to be zero, as shown against $i_0$ in Fig. 6b (the $\kappa_{\perp\mathbf{e}}^{p}/\kappa_{\|\mathbf{e}}^{p} = 0$ curve).

We come to the following conclusions for $v^m = 1/2$ (Fig. 6b): (a) $h < h_c$: When light is weak, the film bends towards the light in the *x-z* plane and bends backward from the light in the *y-z* plane, exhibiting a saddle shape. When light is strong, the film bends towards the light in both the *x-z* and *y-z* plane, exhibiting a bowl shape; (b) $h > h_c$: the film always bends towards the light in both the *x-z* and *y-z* plane, exhibiting a bowl shape. In particular, when $h$ and $i_0$ follow a certain relation such that $\kappa_{\perp \mathbf{e}}^P(h) = 0$ (Fig. 6b), the film bends in an exact cylindrical shape (Fig. 7a). As light curls the film from small deflection to large (not in the category of our small deflection theory), $\varepsilon_{\|\mathbf{e}}^p$ and $\kappa_{\|\mathbf{e}}^p$ develop but $\varepsilon_{\perp \mathbf{e}}^p$ and $\kappa_{\perp \mathbf{e}}^p$ cannot as they are restricted by an increased stiffness for bend in the *y-z* plane, much like that reported in [1].

We illustrate for $v^m = 1/2$ when $\kappa_{\|\mathbf{e}}^p$ is far more important than $\kappa_{\perp \mathbf{e}}^p$. A contour plot of the curvature ratio $\kappa_{\perp \mathbf{e}}^P / \kappa_{\|\mathbf{e}}^P$ in the plane of ($i_0$, *h/d*) is shown in Fig. 6b. The contours of $\left| \kappa_{\perp \mathbf{e}}^p / \kappa_{\|\mathbf{e}}^p \right| \neq 0$ surround the contour of $\kappa_{\perp \mathbf{e}}^p / \kappa_{\|\mathbf{e}}^p = 0$. At $(i_0, h/d) \to (0, 0)$, $0.15 < \left| \kappa_{\perp \mathbf{e}}^p / \kappa_{\|\mathbf{e}}^p \right| < 0.2$. This contour depicts a significantly large parameter region for $\left| \kappa_{\perp \mathbf{e}}^p / \kappa_{\|\mathbf{e}}^p \right| < 0.2$. In [1] $\kappa_{\perp \mathbf{e}}^P$ is not necessarily 0 but may be small compared with $\kappa_{\|\mathbf{e}}^p$ (Fig. 7a). As the bend increases from small deflection to large, the bending stiffness in the *y-z* plane looms so large for $\kappa_{\perp \mathbf{e}}^P$ to develop any further. Experiments are needed to test to what extent $\left| \kappa_{\perp \mathbf{e}}^p / \kappa_{\|\mathbf{e}}^p \right| < 0.2$ can be achieved using the parameters given in Fig. 6b. Hopefully one might observe non-cylindrical bend using parameters for $\left| \kappa_{\perp \mathbf{e}}^P / \kappa_{\|\mathbf{e}}^p \right| > 0.2$!

One might also wonder what $v^m$ is needed to get anticlastic, cylindrical and non-cylindrical bend. Fig. 7b shows a contour plot of the curvature ratio $\left| \kappa_{\perp \mathbf{e}}^P / \kappa_{\|\mathbf{e}}^P \right| = 0.2$ in the plane of ($i_0$, *h/d*) for different Poisson ratios $v^m$. The $\left| \kappa_{\perp \mathbf{e}}^P / \kappa_{\|\mathbf{e}}^p \right| < 0.2$ area for any $v^m$ is surrounded by the $\left| \kappa_{\perp \mathbf{e}}^p / \kappa_{\|\mathbf{e}}^p \right| = 0.2$ curve and both coordinate axes. As $v^m$ decreases, the $\left| \kappa_{\perp \mathbf{e}}^P / \kappa_{\|\mathbf{e}}^p \right| = 0.2$ contours shift leftward and downward, and the $\left| \kappa_{\perp \mathbf{e}}^P / \kappa_{\|\mathbf{e}}^p \right| < 0.2$ area shrinks, say $v^m = 0.4$ in Fig. 7b. As $v^m$ increases, the $\left| \kappa_{\perp \mathbf{e}}^P / \kappa_{\|\mathbf{e}}^p \right| = 0.2$ contours shift rightward and upward, and the $\left| \kappa_{\perp \mathbf{e}}^P / \kappa_{\|\mathbf{e}}^p \right| < 0.2$ area expands, say $v^m = 0.6$ in Fig. 7b. Given a particular sample, hence $h$ and $v^m$ are given,

experimentalists might find an appropriate light intensity from Fig. 7b to observe a desired bending shape.

## 5. Conclusions

We have applied the classical homogenization method in micromechanics to describe effective polydomain deformation. We have analyzed the photo-deformation and bend in polydomain nematic glass. We come to the following conclusions:

1) The popular assumption that axial contraction along polarization and lateral expansion is not always true for polydomain samples. The two in-plane photo-strain components $\varepsilon_{\|\mathbf{e}}^{p}(z)$ and $\varepsilon_{\perp\mathbf{e}}^{p}(z)$ could be either contraction or expansion depending on parameters (Poisson ratio, illumination intensity) and change signs through the material.

2) Neither anticlastic nor cylindrical bend could always be observed ---- observed shape varies according to film thickness, illumination and Poisson ratio. For $\nu^m = 1/2$, a film with thickness $h$ always bends towards the light in the *x-z* plane but is not certain in the *y-z* plane (refer to Fig. 2): (a) $h < h_c$: When light is weak, the film bends backward from the light in the *y-z* plane, exhibiting a saddle shape. When light is strong, the film bends towards the light in both the *x-z* and *y-z* plane, exhibiting a bowl shape; (b) $h > h_c$: the film always bends towards the light in both the *x-z* and *y-z* plane, exhibiting a bowl shape. In particular, when $h$ and $i_0$ follow a certain relation, the film bends in an exact cylindrical shape.

3) A possible explanation for the cylindrical bending shape in [1] is suggested. If bend is predominant in one direction within the small deflection limit, this predominance will be greatly enlarged as the film bends to large deflection ---- due to the bent configuration the bending stiffness in lateral directions ascends to unable further development of bend in lateral directions. A way of reproducing desired bending shapes is given in a parameter plane, also depending on film thickness, illumination and Poisson ratio.

The Sachs limit in calculating polydomain photo-deformation is simple but telling. It brings about a series of rich properties of photo-strain distributions, which result in complicated bending features. From these above we have found ways to reproduce the scenario in experiment [1] by adjusting Poisson ratio, film thickness and illumination intensity, and gave an explanation for how the cylindrical bend happens. One might also be able to observe anticlastic and bowl shapes by adjusting these parameters. We call for further experiments to test how to get non-cylindrical bend.

**Acknowledgements**

The authors acknowledge the support of the National Science Foundation of China (Nos 11461161008, 11272092). We thank Dr. John Simeon Biggins for precious suggestions.

**Figures**

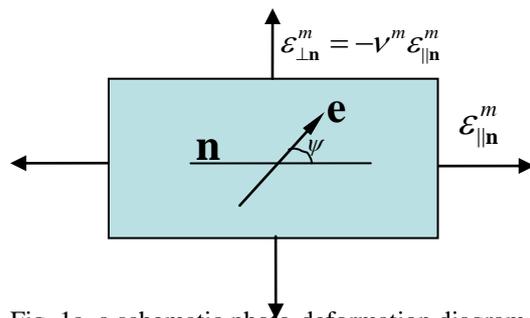

Fig. 1a. a schematic photo-deformation diagram of a monodomain nematic glass illuminated by PUV.

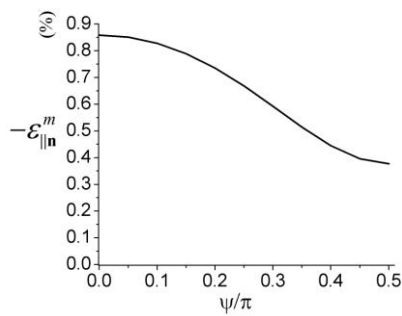

Fig. 1b. surface axial photo-contraction in an LC domain.

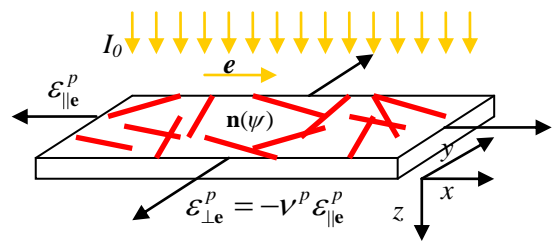

Fig. 2. a schematic diagram of an RVE extracted from a polydomain nematic glass film illuminated by PUV.

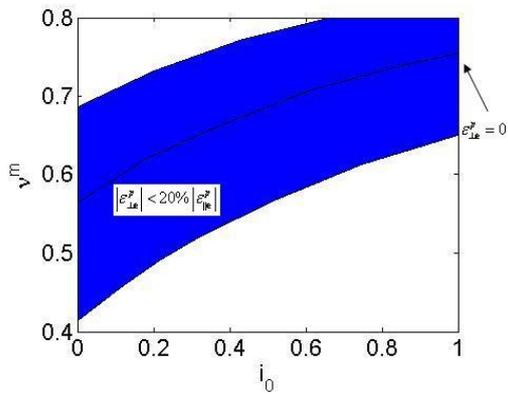

Fig. 3. area for $\left|\varepsilon_{\perp\mathbf{e}}^{p}/\varepsilon_{\|\mathbf{e}}^{p}\right|<20\%$ and curve for $\varepsilon_{\perp\mathbf{e}}^{p}=0$ at the surface of a polydomain nematic glass.

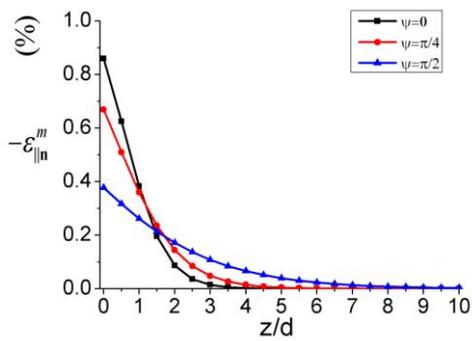

Fig. 4. Axial photo-contraction in an LC domain for different polarization angles.

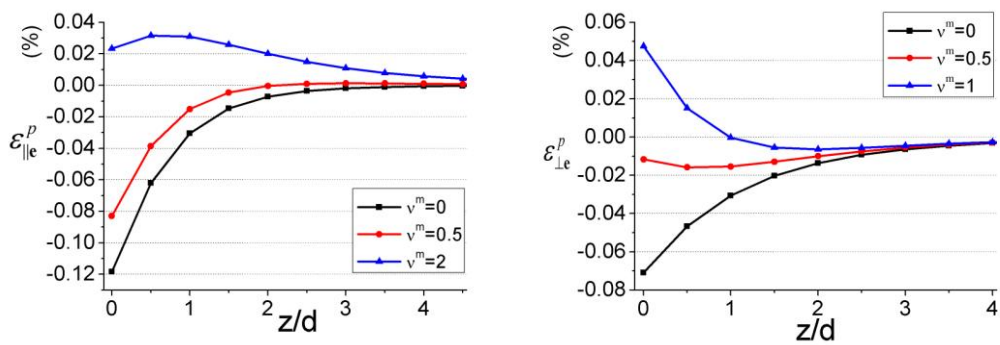

Fig. 5. Axial and lateral photo-strain in a polydomain nematic glass film for different Poisson ratios $\nu^m$ (light intensity $i_0 = 0.1$).

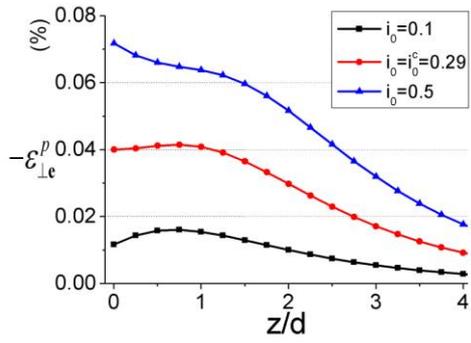

Fig. 6a. lateral photo-strain in a polydomain nematic glass film ($v^m = 1/2$) for different light intensity $i_0$.

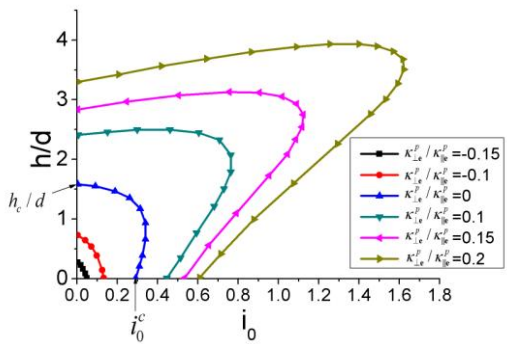

Fig. 6b. contour of curvature ratio $\kappa^p_{\perp e} / \kappa^p_{\parallel e}$ in the plane of ($i_0$, h/d) for a polydomain nematic glass film ($v^m = 1/2$).

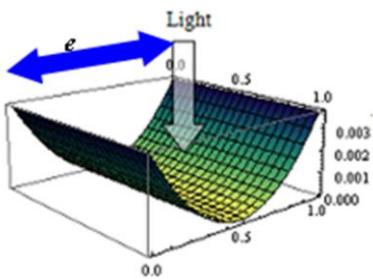

Fig. 7a. a schematic diagram of approximate cylindrical-shaped bend.

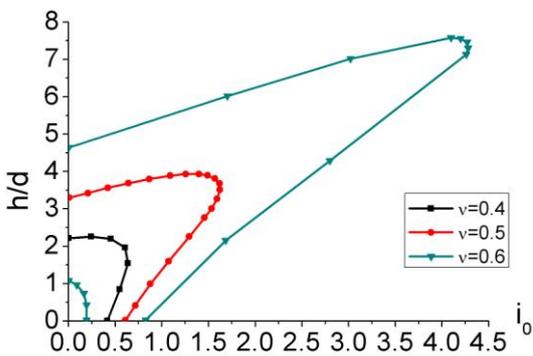

Fig. 7b. contour of curvature ratio $\left|\kappa_{\perp e}^{p} / \kappa_{\|e}^{p}\right| = 0.2$ in the plane of ($i_0$, $h/d$) for different Poisson ratios $\nu^m$.